\documentclass[aps,pra,reprint,amsmath,amssymb,nofootinbib]{revtex4-2}
\usepackage[draft]{changes} 
\usepackage[utf8]{inputenc}
\usepackage[english]{babel}
\usepackage{lipsum}
\usepackage{csquotes}
\usepackage{setspace}
\usepackage{braket}
\usepackage{physics}
\usepackage{booktabs}
\usepackage{soul}
\usepackage{graphicx}
\graphicspath{{Figures/}}
\UseRawInputEncoding
\usepackage{todonotes}

\usepackage{hyperref}
\hypersetup{
    colorlinks=true,
    urlcolor=blue,
    linkcolor=blue,
    citecolor=blue
}

\usepackage{orcidlink}

\begin{document}

\preprint{APS/123-QED}

\title{A measurement-free quantum error correction protocol for superconducting cavities}
\author{Q. Rumman Rahman}
\altaffiliation{Present address: 
Princeton Quantum Initiative, Princeton University, Princeton, NJ 08540, USA}
\author{S. M. Girvin}
\author{D. K. Weiss}
\altaffiliation{Present address: 
D-Wave Quantum Inc., New Haven, CT 06511, USA}
\affiliation{Departments of Applied Physics and Physics, Yale University, New Haven, CT 06511, USA}
\affiliation{Yale Quantum Institute, Yale University, New Haven, CT 06511, USA}

\date{\today}

\begin{abstract}

The realization of large-scale quantum computation will require quantum error correction (QEC). 
One way to reduce the demands placed on QEC is to correct dominant physical errors at a lower level before passing the remaining errors to an outer code.
Recently, Krastanov \textit{et al.} [Nat. Commun. \textbf{12}, 191 (2021)] proposed a measurement-free error-correction protocol based on coherent control of photonic modes that corrects for photon loss. 
Here, we adapt this approach to a circuit quantum electrodynamics (cQED) architecture using three superconducting microwave
cavities coupled by Josephson-junction-based nonlinear elements.
We encode the logical qubit in the lowest-order binomial code of one cavity, while the two other cavities serve as ancillas. Using numerical simulations with experimentally realistic parameters, we demonstrate that our protocol can correct single-photon loss with 99\% fidelity. Furthermore, we optimize the waiting time between successive rounds of error-correction, and show that the logical lifetime can be extended by a factor of 1.2--1.8 relative to the unprotected Fock $\{|0\rangle,|1\rangle\}$ encoding, surpassing break-even performance. Our work highlights the potential of gate-based, measurement-free QEC for fault-tolerant quantum computing.

\end{abstract}

\maketitle

\section{Introduction}

Quantum error correction (QEC) is essential for realizing practical quantum computers and remains one of the outstanding challenges in the field. Traditional fault-tolerant architectures are commonly based on stabilizer codes, such as the surface code \citep{fowler2012surface}, Shor code \citep{shor1995scheme}, Steane code \citep{steane1996error}, and color code \citep{bombin2006topological}. In these schemes, repeated syndrome measurements are used to detect errors, followed by feed-forward recovery operations conditioned on the results. In recent years, experimental demonstrations of measurement-based QEC using circuit quantum electrodynamics (cQED) architectures have shown remarkable progress, decreasing logical errors by increasing the code distance \citep{google2023suppressing,google2025quantum}. However, the physical error rates and control latencies in present devices remain large enough that substantial resources are still required to obtain a practical reduction in the logical error rate.

One promising approach to ease these demands is code
concatenation~\cite{KnillLaflamme1996concatenated},
where each ``physical'' qubit in, e.g., the surface code,
is itself an encoded qubit of an underlying code.
Many such lower-level QEC schemes have been proposed
in a measurement-free form, typically using autonomous
dissipation to remove entropy from the encoded
system~\cite{cohen2014dissipation,kapit2016hardware,
lihm2018implementation,gertler2021protecting,
shirol2025passive}. In these protocols, engineered coupling to lossy modes produces irreversible, non-unitary evolution from the error space back to the code space. Reducing the number of syndrome measurements during QEC is particularly attractive in cQED, where readout and measurement-conditioned feedback can introduce overhead and measurement-induced errors.

Recently, \citet{krastanov2021room} proposed a measurement-free error-correction protocol based on coherent control of photonic modes. Apart from reset operations that remove entropy between subsequent rounds, the protocol is entirely unitary. Here, we adapt this idea to cQED \citep{blais2021circuit}, replacing the Chuang-Leung-Yamamoto (CLY) encoding \citep{chuang1997bosonic} with the lowest-order binomial code \citep{michael2016new}, and thereby reducing the required number of bosonic modes from six to three.

Our choice of encoding is well motivated in superconducting-cavity hardware. Microwave cavity modes have a relatively simple error model, where single-photon loss is the dominant noise process and pure dephasing is strongly suppressed \citep{ofek2016extending}. The lowest-order binomial code directly addresses this model by correcting a single photon-loss event using a finite superposition of Fock states. Indeed, break-even performance with binomial encodings has
already been demonstrated in superconducting-cavity
experiments, using both measurement-based
\citep{ni2023beating,cai2026error} and measurement-free
\citep{ni2025autonomous,sun2026extending,shirol2025passive}
approaches.

In this work, we propose a binomial-code QEC protocol which does not use syndrome measurements.  We do so by leveraging the inherent strengths of cQED systems including strong, tunable interactions between microwave modes \cite{chapman2022high, lu2023high} and millisecond-scale coherence times \citep{bland2025millisecond, ganjam2024surpassing, milul2023superconducting, reagor2016quantum}. Our architecture consists of three microwave cavities connected by Josephson-junction-based nonlinear elements, with auxiliary transmons coupled to each cavity for local operations, state preparation, and readout. 

A key feature of our architecture is that the majority of the error-correction operations are generated by parametrically activated cavity-cavity exchange interactions \cite{chapman2022high, lu2023high}, rather than by transmon-mediated dispersive control. This distinction is important because many bosonic cQED protocols rely on cavity and transmon drives for most non-Gaussian operations. In such schemes, the dispersive coupling can limit the relevant gate speed, while transmon relaxation and dephasing can contribute substantially to the gate infidelity \cite{pietikainen2024strategies}. By contrast, our architecture is designed to suppress the impact of auxiliary-transmon decoherence and to focus the error-correction task on the dominant cavity photon-loss channel.

Within this platform, we design a protocol consisting of a series of numerically optimized gates realized with the gradient ascent pulse engineering algorithm (GRAPE) \citep{khaneja2005optimal}. Our protocol targets the dominant single-photon-loss event while also compensating the deterministic no-jump back-action associated with the absence of photon loss. The latter is necessary because unlike fixed-total-photon-number encodings such as the CLY code, the lowest-order binomial code contains unequal photon-number components that acquire different weights under the no-jump evolution. Using numerical simulations with experimentally realistic parameters, we demonstrate that a single round of error correction can achieve 99\% correction fidelity, with transmon decoherence contributing only $\approx10\%$ of the single-round infidelity budget. Moreover, by optimizing the waiting time between successive rounds of error correction, we obtain logical lifetime enhancements of $1.2$--$1.8$ relative to the unprotected Fock $\{|0\rangle,|1\rangle\}$ encoding, for cavity lifetimes ranging from $0.5$ to $2\,\mathrm{ms}$.

The paper is organized as follows. In Sec.~\ref{sec:error_correction}, we introduce the error correction strategy and outline the supporting architecture. In Sec.~\ref{sec:gates}, we discuss the required gate operations and detail their implementation through the system Hamiltonian. In Sec.~\ref{sec:results}, we present numerical simulations of the measurement-free protocol, benchmarking its potential for robust QEC beyond break-even.

\section{Binomial Code QEC}
\label{sec:error_correction}
The binomial code is a single-mode code whose codewords written in the photon number (Fock state) basis are
\begin{align}
     |0_{\textnormal{L}}\rangle &= \frac{|0\rangle + |4\rangle}{\sqrt{2}}\\
     |1_{\textnormal{L}}\rangle &= |2\rangle.
\end{align}
We first discuss the typical measurement-based approach to QEC with the binomial code. We then outline how a measurement-free protocol could work in principle using only two modes. Finally, we turn to our measurement-free proposal which requires three modes, and discuss the resulting QEC protocol in detail.

\subsection{The Measurement-Based Approach}

Traditional measurement-based error correction with the binomial code relies on two components: a storage mode and an ancilla qubit (e.g., a transmon) \citep{michael2016new, hu2019quantum, ni2023beating}. The storage mode encodes the logical codewords, while the ancilla is responsible for detecting and correcting errors.
We denote an arbitrary state in this setup as \(\ket{n_\textrm{S}\, n_\textrm{A}}\), where \(n_\textrm{S}\)  represents the photon number in the storage mode and \(n_\textrm{A}\) is the excitation number of the ancilla qubit. The logical codewords including the state of the ancilla are \( |0_{\textnormal{L}}\rangle = \frac{|0\,0\rangle + |4\,0\rangle}{\sqrt{2}} \) and \( |1_{\textnormal{L}}\rangle = |2\,0\rangle \).
When a single-photon loss occurs in the storage mode, the logical states are mapped onto the error space, spanned by \( |0_{\textnormal{E}}\rangle = |3\,0\rangle \) and \( |1_{\textnormal{E}}\rangle = |1\,0\rangle \). This error syndrome can be detected via a parity operation \cite{sun_tracking_2014}, which performs the mapping  $|j\,0\rangle\rightarrow|j\,(j \mod{2})\rangle$. If the ancillary mode is measured in \( \ket{1} \), the ancilla is reset and a recovery operation is performed to restore the logical states  \(|1\, 0\rangle \rightarrow |2\, 0\rangle, |3\, 0\rangle \rightarrow \frac{|0\,0\rangle + |4\,0\rangle}{\sqrt{2}} \). Conversely, if the ancilla is not in the excited state, no recovery is applied, leaving the storage mode unchanged. For simplicity of presentation here and throughout the remainder of this section, we neglect the deterministic no-jump back-action associated with conditional evolution in the absence of photon loss \cite{michael2016new}. This effect is explicitly accounted for  in Sec.~\ref{sec:gates}.

\subsection{Measurement-Free QEC with Two Modes}
\label{sec:two_cavities}
In principle, a two-mode system consisting of a storage mode and an ancilla could realize a measurement-free error-correction cycle, provided that universal control and unconditional ancilla reset are available. Rather than mapping the parity of the storage mode onto the ancilla and measuring it, the measurement-free cycle begins by preparing the ancilla in the excited state $|1\rangle$. A joint unitary on the storage and ancilla modes then implements the desired conditional recovery: in the no-error subspace, it leaves the codewords unchanged,
$|2\,1\rangle\rightarrow |2\,1\rangle$ and
$\frac{1}{\sqrt{2}}(|0\rangle+|4\rangle)|1\rangle\rightarrow
\frac{1}{\sqrt{2}}(|0\rangle+|4\rangle)|1\rangle$;
in the error subspace, it restores the codewords by transferring the ancilla excitation into the storage,
\begin{align}
|1\,1\rangle &\rightarrow |2\,0\rangle,\\
|3\,1\rangle &\rightarrow \frac{1}{\sqrt{2}}(|0\rangle+|4\rangle)|0\rangle.
\end{align}
Unlike the measurement-based approach, this correction step is a coherent conditional unitary and therefore eliminates the need for ancilla measurement and real-time feedback. At the end of the cycle, an unconditional reset of the ancilla removes entropy and prepares the system for the next round of correction. This type of two-mode measurement-free strategy has recently been demonstrated beyond break-even using a transmon qubit as the ancilla \cite{ni2025autonomous}.

Here, we pursue a complementary approach. Rather than realizing the recovery through direct storage--ancilla control, we use a three-cavity architecture in which the central recovery operations are implemented with parametrically activated cavity--cavity exchange interactions. Although these interactions are generated by Josephson-junction-based nonlinear couplers, the coupler modes are not directly excited, and enter only through the effective exchange Hamiltonian. This reduces reliance on dispersive control through the auxiliary transmons, whose relaxation and dephasing can otherwise contribute significantly to the infidelity. We now describe the resulting correction cycle at the level of state mappings, before discussing the Hamiltonian implementation in Sec.~\ref{sec:gates}.

\renewcommand{\figurename}{FIG.}
\begin{figure}[!t]
    \centering
    \includegraphics[width=\linewidth]{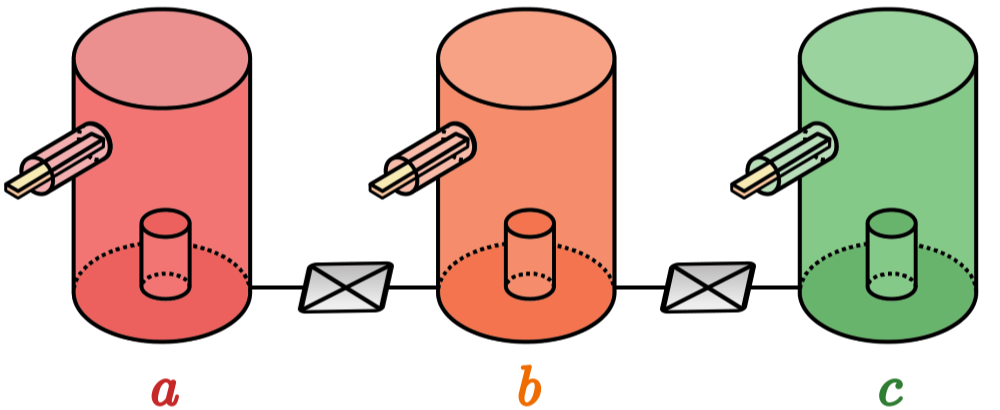}
    \caption{\textbf{Schematic of the proposed architecture.} The setup comprises three microwave cavities, labeled $a$ (red), $b$ (orange), and $c$ (green). Each cavity is dispersively coupled to a transmon qubit (yellow), enabling control and readout of the cavity state. Additionally, the cavities are pairwise coupled via Josephson-junction-based nonlinear elements (grey).
\label{fig:figure1}}
\end{figure}

\subsection{Measurement-Free QEC with Three Cavities}
\label{subsec:three_cavities}

Fig.~\ref{fig:figure1} illustrates our proposed setup comprising three microwave cavities, labeled $a$, $b$, and $c$. Cavity $b$ functions as the storage mode, while cavities $a$ and $c$ serve as ancillas. Each cavity is dispersively coupled to a transmon qubit \citep{koch2007charge}, which enables state preparation and readout of the corresponding cavity, as well as the implementation of single-cavity gates used in the error-correction protocol.

Cavity-cavity interactions are mediated by two Josephson-junction-based coupling elements \citep{gao2019entanglement, lu2023high, chapman2022high}: one between cavities $a$ and $b$, and another between cavities $b$ and $c$. Specifically, our protocol utilizes a two-to-one exchange coupling \cite{vrajitoarea2020quantum} between cavities $a$ and $b$ and a one-to-one beamsplitter coupling \citep{gao2019entanglement, lu2023high, chapman2022high} between cavities $b$ and $c$, both of which can be enabled by four-wave mixing elements. Together, the single-cavity gates, implemented using cavity and transmon drives, and the multi-cavity gates, implemented using coupler drives, form the basis of the error-correction protocol. 

Denoting a three-cavity Fock state by $\ket{n_a\,n_b\,n_c}$, where $n_i$ is the occupation number of cavity $i$, the logical codewords are
\begin{equation}
\begin{aligned}
|0_{\textnormal{L}}\rangle &= \frac{|0\,0\,0\rangle + |0\,4\,0\rangle}{\sqrt{2}}, \\
|1_{\textnormal{L}}\rangle &= |0\,2\,0\rangle.
\end{aligned}
\end{equation}
Cavity $b$ stores the binomial codewords, while cavities $a$ and $c$ are initialized in their vacuum states. A single-photon loss in $b$ maps the codewords into the error space:
\begin{equation}
\begin{aligned}
\\[-1em]
|0_{\textnormal{L}}\rangle \rightarrow|0_{\textnormal{E}}\rangle &= |0\,3\,0\rangle, \\
|1_{\textnormal{L}}\rangle \rightarrow
|1_{\textnormal{E}}\rangle &= |0\,1\,0\rangle.
\end{aligned}
\end{equation}
Correcting this error therefore involves pumping a photon back into $b$, as well as restoring the superposition present in the $|0_{\rm{L}}\rangle$ state.

The protocol begins with a single-cavity ``pumping'' gate, where one photon is pumped into cavity $c$ by means of an ancilla qubit.
Following this operation, the ``error'' and ``no-error'' states are transformed as:
\begin{equation}
\begin{array}{c c c}
\textbf{Error} & & \textbf{No-Error} \\
\begin{aligned}
\\[-0.5em]
|0\,3\,0\rangle&\rightarrow |0\,3\,1\rangle, \\[0.5em]
|0\,1\,0\rangle &\rightarrow |0\,1\,1\rangle,
\end{aligned}
& \text{} &
\begin{aligned}
\\[-1em]
\frac{|0\,0\,1\rangle + |0\,4\,1\rangle}{\sqrt{2}}&\rightarrow \frac{|0\,0\,1\rangle + |0\,4\,1\rangle}{\sqrt{2}}, \\
|0\,2\,1\rangle&\rightarrow |0\,2\,1\rangle.
\end{aligned}
\end{array}
\end{equation}
The pumped photon serves as a flag for subsequent steps, allowing the protocol to differentiate between the error and no-error cases.

Next, we apply a multi-cavity ``conditional pumping'' gate, which coherently transfers the flag photon from cavity $c$ to cavity $b$ via the one-to-one beamsplitter interaction, but only when the storage mode lies in the single-loss error space. In the error branch, this transfer restores the missing photon and returns cavity $c$ to vacuum; in the no-error branch, the storage state is left unchanged and the flag photon remains in $c$. Explicitly, the states are mapped to
\begin{equation}
\begin{array}{c c c}
\textbf{Error} & & \textbf{No-Error} \\
\begin{aligned}
\\[-0.5em]
|0\,3\,1\rangle &\rightarrow |0\,4\,0\rangle, \\[0.7EM]
|0\,1\,1\rangle &\rightarrow |0\,2\,0\rangle,
\end{aligned}
& \text{} &
\begin{aligned}
\\[-1em]
\frac{|0\,0\,1\rangle + |0\,4\,1\rangle}{\sqrt{2}} &\rightarrow \frac{|0\,0\,1\rangle + |0\,4\,1\rangle}{\sqrt{2}}, \\
|0\,2\,1\rangle&\rightarrow |0\,2\,1\rangle.
\end{aligned}
\end{array}
\end{equation}

With the photon loss reversed, we next recover the superposition of 0 and 4 photons in $|0_{\textnormal{L}}\rangle$. This recovery is performed in cavity $a$, so the error branch must first be routed from cavity $b$ to cavity $a$ without disturbing the no-error branch. The multi-cavity ``conditional routing" gate accomplishes this by acting conditioned on the occupation of cavity $c$. When cavity $c$ is in vacuum, which identifies the error branch, the gate routes population from $b$ to $a$ via the two-to-one exchange coupling between the two cavities. When the flag photon remains in cavity $c$, corresponding to the no-error branch, the gate acts as the identity and leaves the state unchanged:
\begin{equation}
\begin{array}{c c c}
\textbf{Error} & & \textbf{No-Error} \\
\begin{aligned}
\\[-0.4em]
|0\,4\,0\rangle&\rightarrow |2\,0\,0\rangle, \\[0.7EM]
|0\,2\,0\rangle &\rightarrow |1\,0\,0\rangle,
\end{aligned}
& \text{} &
\begin{aligned}
\\[-1em]
\frac{|0\,0\,1\rangle + |0\,4\,1\rangle}{\sqrt{2}}&\rightarrow \frac{|0\,0\,1\rangle + |0\,4\,1\rangle}{\sqrt{2}}, \\
|0\,2\,1\rangle&\rightarrow |0\,2\,1\rangle.
\end{aligned}
\end{array}
\end{equation}
The conditional routing gate must also act as the identity on the states $|2\,4\,1\rangle$, $|2\,2\,1\rangle$, and $|2\,0\,1\rangle$, a requirement that follows from the subsequent action of the superposition gate (discussed below).

We are now ready to recover the superposition of $|0_{\rm{L}}\rangle$. With the photons corresponding to the error state having been mapped from cavity $b$ to cavity $a$, we first apply a single-cavity ``superposition gate", which performs the mapping: \( |2\rangle_{a} \rightarrow \frac{|0\rangle_{a} + |2\rangle_{a}}{\sqrt{2}} \) and \( |1\rangle_{a} \rightarrow |1\rangle_{a} \). In the error case, this restores the superposition $|2\,0\,0\rangle \rightarrow \frac{|0\,0\,0\rangle + |2\,0\,0\rangle}{\sqrt{2}}$, and leaves $|1\,0\,0\rangle$  unchanged. Applying the conditional routing gate one more time then returns the states to $b$, fully recovering the original codewords: 
\begin{equation}
\begin{array}{c c c}
\textbf{Error} \\
\begin{aligned}
\\[-1em]
\frac{|0\,0\,0\rangle + |2\,0\,0\rangle}{\sqrt{2}} &\rightarrow \frac{|0\,0\,0\rangle + |0\,4\,0\rangle}{\sqrt{2}}, \\
|1\,0\,0\rangle &\rightarrow |0\,2\,0\rangle.
\end{aligned}
\end{array}
\end{equation}

\renewcommand{\figurename}{FIG.}
\begin{figure*}[!t]
\centering
\includegraphics[width=\textwidth]{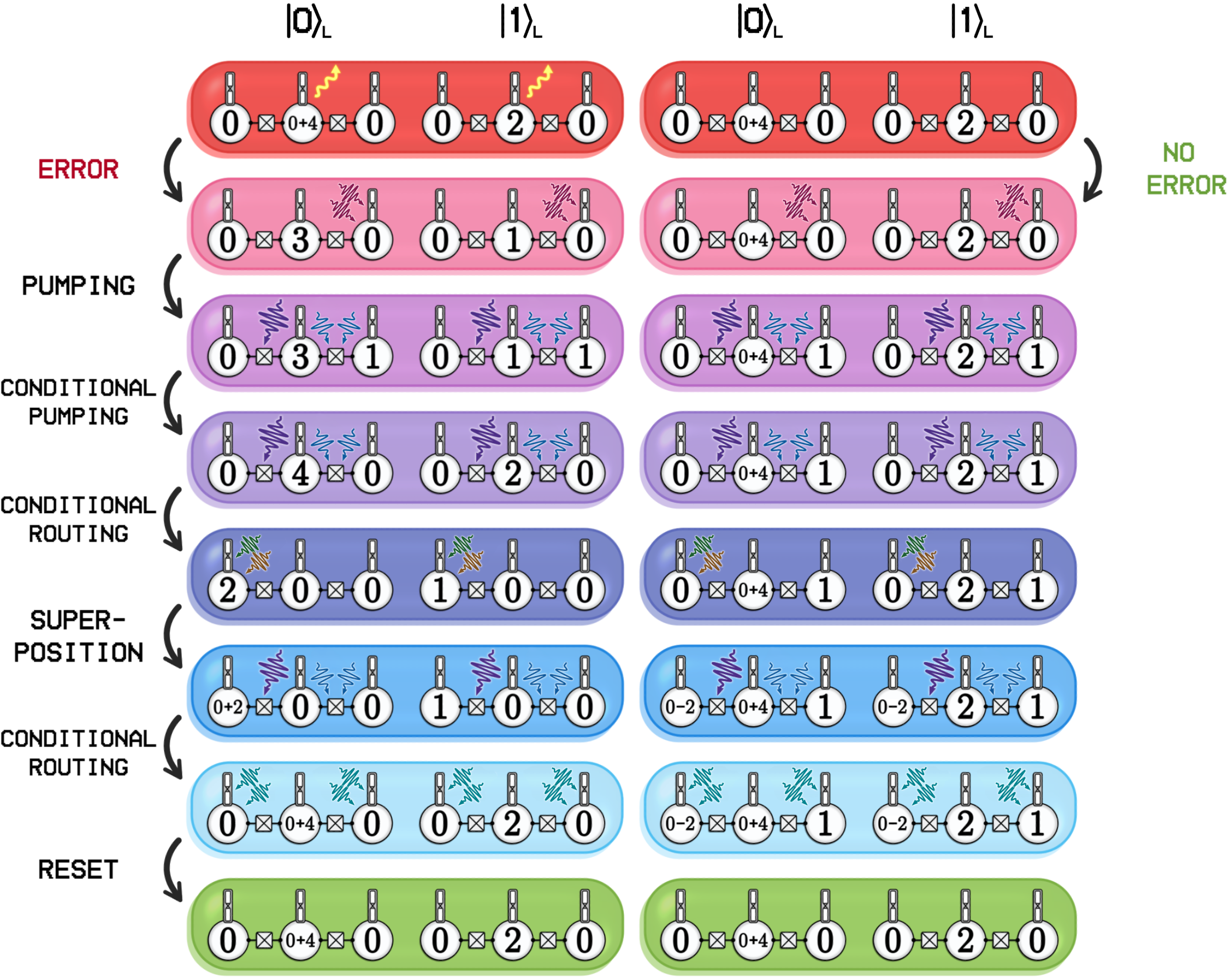}
\caption{\textbf{The measurement-free QEC protocol.} Left (right) column shows the evolution of the logical basis states in the error (no-error) branch. Note that for clarity, this schematic ignores the deterministic no-jump back-action associated with the absence of photon loss. The protocol begins with a pumping operation that prepares a flag photon in cavity $c$ (pink arrows indicate drives on cavity $c$ and its auxiliary transmon). Next, the conditional pumping gate is applied using coupler drives (purple arrow on the $a$--$b$ coupler and blue arrow on the $b$--$c$ coupler): in the error branch this transfers the flag from $c$ to $b$ and restores the lost photon, while in the no-error branch the flag remains in $c$ and cavity $b$ is left unchanged. The conditional routing gate, implemented by driving the same couplers (purple arrow on $a$--$b$ and blue arrow on $b$--$c$), routes population between cavities $b$ and $a$ conditioned on the flag photon, isolating the error branch for recovery. A single-cavity superposition gate on cavity $a$ (brown arrow on $a$ and green arrow on its transmon) reconstructs the $|0_{\mathrm{L}}\rangle$ superposition; in the no-error branch this produces a deterministic ancillary excitation that is removed at the end of the cycle. Applying conditional routing again returns the logical state to cavity $b$. Finally, an unconditional reset returns cavities $a$ and $c$ to vacuum (pale blue arrows on cavities $a,c$ and their transmons), preparing the system for the next round.
\label{fig:figure2}}
\end{figure*}

\noindent The no-error branch requires one additional consideration. Before the superposition gate is applied, cavity $a$ is in $\ket{0}_a$. The recovery condition fixes the action of the superposition gate on the error states as described above, but it does not uniquely determine the action on $\ket{0}_a$. We choose the gate to be the minimal unitary extension acting within the $\{\ket{0}_a,\ket{2}_a\}$ subspace,
\begin{equation}
\begin{aligned}
\ket{2}_a &\rightarrow \frac{\ket{0}_a+\ket{2}_a}{\sqrt{2}}, \\
\ket{0}_a &\rightarrow \frac{\ket{0}_a-\ket{2}_a}{\sqrt{2}}.
\end{aligned}
\end{equation}
As a result, the states \(\ket{0\,4\,1}\), \(\ket{0\,2\,1}\), and \(\ket{0\,0\,1}\) associated with the no-error branch are mapped to 
\begin{equation}
\begin{array}{c c c}
\textbf{No-Error} \\
\begin{aligned}
\\[-1em]
|0\,4\,1\rangle &\rightarrow \left(\frac{|0\rangle - |2\rangle}{\sqrt{2}}\right)\otimes|41\rangle,\\
|0\,2\,1\rangle &\rightarrow \left(\frac{|0\rangle - |2\rangle}{\sqrt{2}}\right)\otimes|21\rangle,\\
|0\,0\,1\rangle &\rightarrow \left(\frac{|0\rangle - |2\rangle}{\sqrt{2}}\right)\otimes|01\rangle.
\end{aligned}
\end{array}
\end{equation}
This explains why the final conditional-routing gate must act as the identity on \(\ket{0\,4\,1}\), \(\ket{0\,2\,1}\), \(\ket{0\,0\,1}\), \(\ket{2\,4\,1}\), \(\ket{2\,2\,1}\), and \(\ket{2\,0\,1}\): routing the state $\frac{\ket{0}_{a} - \ket{2}_{a}}{\sqrt{2}}$ from cavity $a$ to cavity $b$ would corrupt the logical information stored in $b$. At this stage, the error branch has been restored to the logical code space, while the no-error branch contains known residual excitations in cavities $a$ and $c$. The final step of the protocol is therefore an unconditional reset of both cavities, returning them to vacuum in preparation for the next round of error correction. It is this non-unitary reset which removes the entropy associated with the error.

In summary, the protocol starts with the initialization of cavity $c$ with a flag photon. If a single-photon loss occurs in $b$, the flag is transferred from $c$ to $b$ to restore the lost photon. The population in $b$ is then conditionally routed to $a$, where the superposition can be recovered. Once the recovery is completed, the state is routed back to $b$, fully recovering the original codewords. The above steps evolve under measurement-free, unitary dynamics. Finally, an unconditional reset of cavities $a$ and $c$ prepares the system for the next round of error correction. 
Linearity ensures that the protocol can correct single-photon loss in any arbitrary superposition $|\psi\rangle = \alpha|0_{\textnormal{L}}\rangle + \beta|1_{\textnormal{L}}\rangle$. The evolution of the basis states $|0_{\textnormal{L}}\rangle$ and $|1_{\textnormal{L}}\rangle$ during the protocol is mapped in Fig. \ref{fig:figure2}.

\section{Gates}
\label{sec:gates}

We now turn to the physical implementation of the gates outlined above, beginning with the two multi-cavity gates---conditional pumping and conditional routing---before discussing the single-cavity gates.

\subsection{Multi-Cavity Gates}
Inspired by Krastanov \textit{et al.}~\cite{krastanov2021room} and taking advantage of the advanced control capabilities of circuit QED, we consider an exchange-only effective Hamiltonian capable of generating the state-to-state mappings required for the multi-cavity gates. In the interaction picture, this Hamiltonian takes the form
\begin{align}
\label{eq:krastanov_ham}
H(t)
=
\Omega_1(t)\,\hat{a}\hat{b}^{\dagger 2}
+
\Omega_2(t)\,\hat{b}^{\dagger}\hat{c}
+
\mathrm{H.c.},
\end{align}
where $\hat{a}\hat{b}^{\dagger 2}$ implements a two-to-one exchange between cavities $a$ and $b$, and $\hat{b}^{\dagger}\hat{c}$ implements a beamsplitter interaction mediating photon exchange between cavities $b$ and $c$. Equation~\eqref{eq:krastanov_ham} should be understood as an effective Hamiltonian obtained after transforming to the rotating frame and retaining only resonant terms under the rotating-wave approximation. In our simulations, we neglect residual cavity nonlinearities arising from the coupling elements.\footnote{In a physical implementation, these include self-Kerr terms $K_j\hat n_j(\hat n_j-1)$ and cross-Kerr terms $\chi_{jk}\hat n_j\hat n_k$. These terms introduce photon-number-dependent phases and detunings, which can in principle be calibrated or incorporated directly into the optimal-control design.}

Similar to the Hamiltonian described in Krastanov {\it et al.}, our multi-cavity Hamiltonian includes a time-dependent modulation of the beamsplitter interaction, with the amplitude given by $\Omega_2(t)$. Whereas Krastanov \textit{et al.} assume a constant coefficient for the \( \hat{a}\hat{b}^\dagger{}^2 \) interaction (due to their use of a \(\chi^{(2)}\) nonlinearity) \cite{krastanov2021room}, we derive our nonlinearity from four-wave mixing processes in Josephson-junction-based elements. This allows us to treat the coefficient $\Omega_1(t)$ as explicitly time-dependent, offering an important additional degree of freedom for gate design.

Note that the operators \( \hat{a}\hat{b}^\dagger{}^2 \) and \( \hat{b}^\dagger \hat{c} \) both commute with \( 2\hat{n}_a + \hat{n}_b + \hat{n}_c \), implying that the Hamiltonian preserves the value of this observable. As a result, subspaces defined by different integer values of \( 2\hat{n}_a + \hat{n}_b + \hat{n}_c \) do not mix under the multi-cavity evolution \cite{krastanov2021room}. Importantly, \( \hat{a}\hat{b}^\dagger{}^2 \) and \( \hat{b}^\dagger \hat{c} \) do not commute with each other, and repeated commutators of these operators do not commute with the operators themselves. This guarantees that the operators can generate a complete Lie algebra within each \( 2\hat{n}_a + \hat{n}_b + \hat{n}_c \) subspace \citep{krastanov2021room, niu2018qudit, bartlett2002efficient, jacobs2007engineering, krastanov2015universal}. In other words, the Hamiltonian in Eq.~\eqref{eq:krastanov_ham} can implement any unitary $U(t)\ket{n_a n_b n_c} = \ket{n_a' n_b' n_c'}$ satisfying $2n_a + n_b + n_c = 2n_a' + n_b' + n_c'$. This is the core principle underlying our multi-cavity operations. The state-to-state mappings defined in the previous section are possible because each gate operates within a subspace defined by a constant value of $2n_a + n_b + n_c$, allowing targeted evolutions within that subspace.

\begin{figure}[!t]
    \centering
    \includegraphics[width=\linewidth]{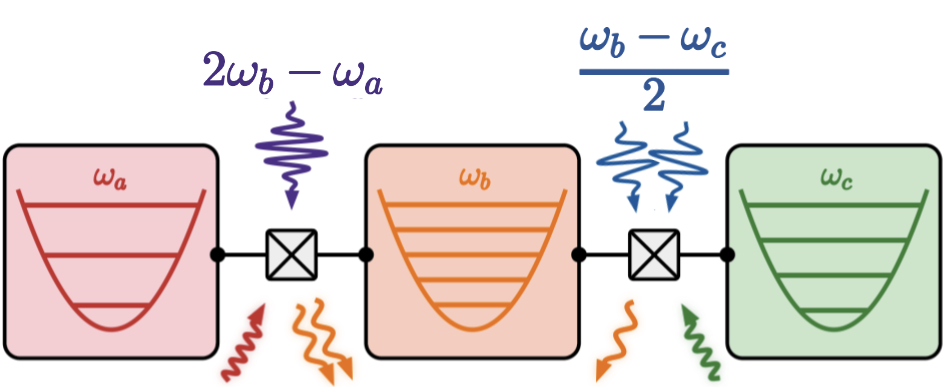}
    \caption{\textbf{Driving fields for the multi-cavity gates.} The conditional-pumping and conditional-routing gates are implemented using microwave drives on the two nonlinear couplers, which activate two multi-cavity interactions. A drive on the $a$-$b$ coupler at $2\omega_b-\omega_a$ (purple) activates the two-to-one exchange interaction $\hat{a}\hat{b}^{\dagger 2}+{\rm H.c.}$, assuming the coupler supports four-wave mixing. A drive on the $b$-$c$ coupler at $(\omega_b-\omega_c)/2$ (blue) activates the beamsplitter interaction $\hat{b}^\dagger \hat{c}+{\rm H.c.}$, again assuming four-wave mixing. 
\label{fig:figure3}
}
\end{figure}

The \(\hat{a}\hat{b}^\dagger{}^2 \) and \( \hat{b}^\dagger \hat{c} \) interactions can be realized using the tunable nonlinearities of the Josephson-junction-based coupling elements \citep{frattini20173, chapman2022high, pietikainen2024strategies, pietikainen2022controlled}. In the lab frame, a time-dependent drive applied to either coupler can be written as
\begin{align}\label{eq:6}
H_{\text{drive}}(t) = \Omega(t) e^{-i \omega_d t} \hat{s}^\dagger + \text{H.c.},
\end{align}
\noindent where $\hat{s}$ is the bosonic annihilation operator of the driven coupling element, $\omega_d$ is the drive frequency, $\Omega(t)$ is the complex drive envelope and we have made the rotating-wave approximation (RWA). Assuming that both couplers are four-wave-mixing elements, a drive on the $a$--$b$ coupler at $\omega_d = 2\omega_b-\omega_a$ resonantly activates the two-to-one exchange interaction $\hat{a}\hat{b}^{\dagger 2}+\mathrm{H.c.}$, while a drive on the $b$--$c$ coupler at $\omega_d = (\omega_b-\omega_c)/2$ activates the beamsplitter interaction $\hat{b}^\dagger \hat{c}+\mathrm{H.c.}$ \cite{lu2023high, pietikainen2024strategies}, see Fig.~\ref{fig:figure3}.
With the couplers driven in this way, we move to the rotating frame of the cavity modes and apply the rotating-wave approximation. After eliminating the coupler modes and neglecting residual self- and cross-Kerr nonlinearities of the cavities, the effective Hamiltonian reduces to Eq.~\eqref{eq:krastanov_ham}. Finally, we use GRAPE to shape the control envelopes $\Omega_1(t)$ and $\Omega_2(t)$ so as to realize the desired state-to-state mappings described in Sec.~\ref{subsec:three_cavities}.

\renewcommand{\figurename}{FIG.}

\renewcommand{\figurename}{FIG.}
\begin{figure*}[!t]
    \centering
    \includegraphics[width=\textwidth]{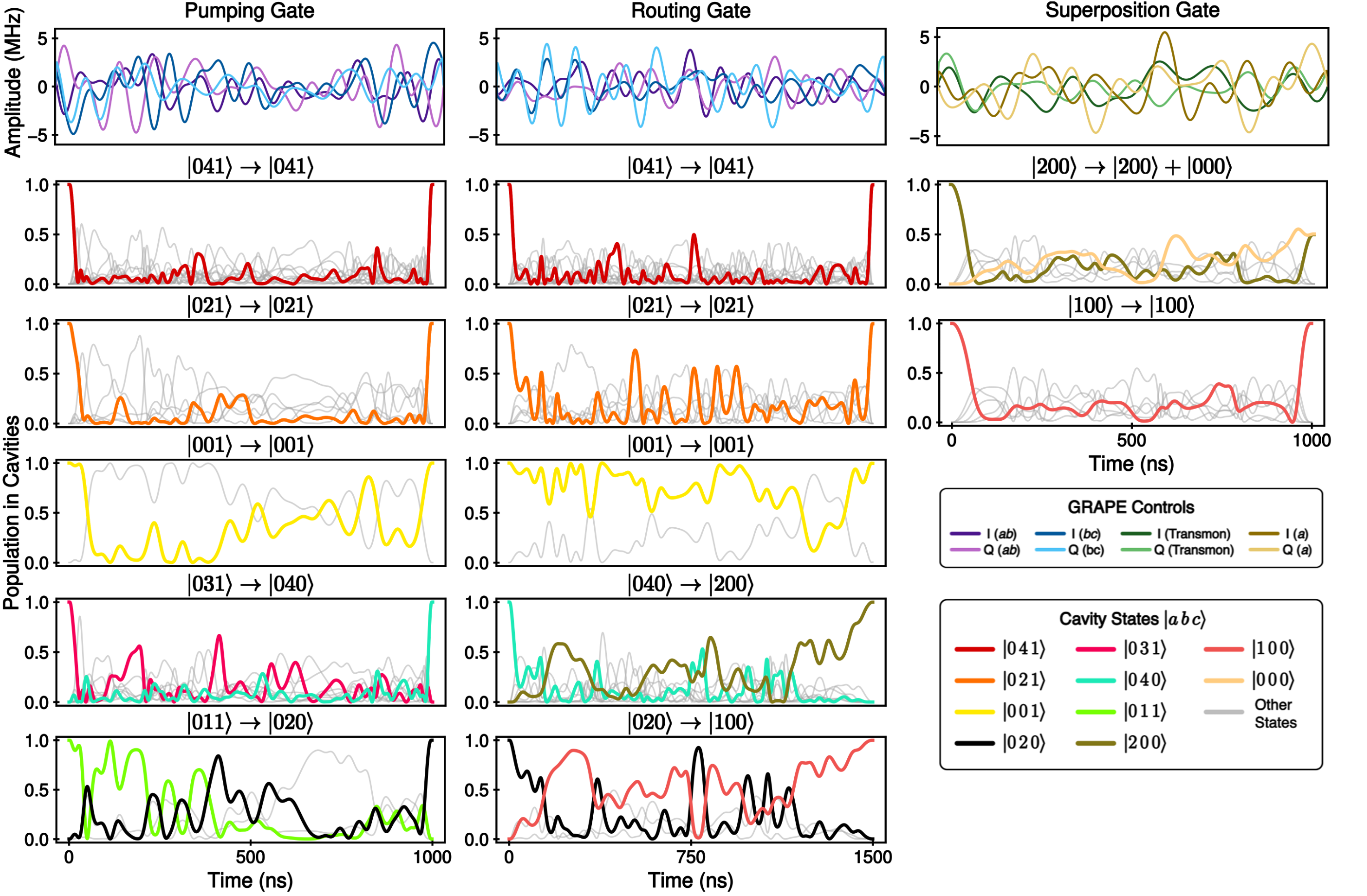}
    \caption{\textbf{Control pulses and state-to-state mappings.} We use GRAPE optimization to obtain relatively smooth pulses for implementing the conditional pumping, conditional routing, and superposition gates, as shown in the top row. The conditional pumping and conditional routing gates require drives applied to both the $a$-to-$b$ coupler (Drive $ab$) and the $b$-to-$c$ coupler (Drive $bc$). The superposition gate is implemented by driving cavity $a$ and the transmon coupled to it. The relevant states undergo highly nontrivial evolutions during each gate, consistent with optimal-control-style pulses, see the remaining panels. Nevertheless, during the two multi-cavity gates, each state can only exchange population with other states with the same value of $2n_a + n_b + n_c$.
    \label{fig:figure4}}
\end{figure*}

\subsection{Single-Cavity Gates}
We now turn to the complementary single-cavity gates required for pumping, state preparation, superposition, and unconditional reset.  In our protocol, the superposition gate can be implemented by simultaneously driving cavity $a$ and the transmon coupled to it. In the rotating frame and truncating to the two lowest transmon levels, the dispersive interaction Hamiltonian is given by 
\begin{equation}\label{eq:9}
    H_{\text{}}(t)
    =
    \frac{\chi}{2}\,\hat{a}^{\dagger}\hat{a}\,\hat{\sigma}_{z}
    +
    \frac{A(t)}{2}\,\hat{\sigma}_{x}
    +
    \epsilon(t)\,\hat{a}+\epsilon^{*}(t)\,\hat{a}^{\dagger}
\end{equation}
\noindent where $\chi$ is the dispersive shift between the cavity and transmon, $A(t)$ is the transmon drive, and $\epsilon(t)$ is the cavity drive. Once again, we use GRAPE to design the amplitude of the driving fields (see Sec.~\ref{subsec:GRAPE}).

In the simulations below, state preparation and reset are treated as ideal operations. Experimentally, these steps would need to be implemented with sufficiently high fidelity. Fock-state preparation in cavities coupled to ancilla transmons has already been demonstrated in previous works \citep{heeres2015cavity, reagor2016quantum, chou2023demonstrating}, providing a viable method for realizing the pumping gate. Additionally, the unconditional reset of cavities \(a\) and \(c\) can be performed by e.g. directly swapping the excitations to a lossy mode \citep{gao2018programmable, axline2018demand}.

\subsection{GRAPE Optimization}
\label{subsec:GRAPE}

With the Hamiltonians for the conditional pumping, conditional routing, and superposition gate defined above, we compute the time-dependent control fields using closed-system GRAPE. For the two multi-cavity gates, the optimization is simplified by a symmetry of the exchange Hamiltonian. As discussed above, the interactions $\hat a\hat b^{\dagger 2}$ and $\hat b^\dagger \hat c$ conserve
$2\hat n_a+\hat n_b+\hat n_c,$
so the dynamics decompose into independent invariant sectors labeled by fixed values of $2n_a+n_b+n_c$. The GRAPE optimization can therefore be restricted to the relevant sectors, rather than performed over the full three-cavity Hilbert space.

For each gate unitary $U$, we achieve an average coherent state-transfer fidelity,  
\begin{align}
F_{\mathrm{coh}}(U)
=
\left|
\frac{1}{N_U}
\sum_{k=1}^{N_U}
\langle\psi_{\mathrm{tgt}}^{(k)}|
U|\psi_{\mathrm{init}}^{(k)}\rangle
\right|^2,
\end{align}
of at least 0.995.  Here \(N_U\) is the number of initial states \(\{|\psi_{\text{init}}^{(k)}\rangle\}\) included in the optimization (recall that only a subset of logical states is relevant for each gate), and \(\{|\psi_{\text{tgt}}^{(k)}\rangle\}\) are the corresponding post-gate target states.

Importantly, because the GRAPE optimization is closed-system, it does not automatically capture the deterministic no-jump back-action on cavity $b$. This back-action must be included because $|0_{L}'\rangle$ is a superposition of Fock states with different photon numbers, so conditioning on the absence of a photon-loss event preferentially weights the lower-photon-number component. To make our pulses robust to this effect to leading order, we apply a pre-rotation to the target states of the no-error branch. To see how this is done, first recall that following the pumping operation along the no-error branch, the three-cavity state is 
\( |\psi\rangle = u\,|0_{\text{L}}'\rangle + v\,|1_{\text{L}}'\rangle \),
where 
\( |0_{\textnormal{L}}'\rangle = \frac{|0\ 0\ 1\rangle + |0\ 4\ 1\rangle}{\sqrt{2}}\) and \( |1_{\textnormal{L}}'\rangle = |0\ 2\ 1\rangle\). The evolution conditioned
on no photon loss in the storage cavity $b$ during a gate of duration \(\delta t\) is governed by the deterministic, non-unitary Kraus operator
\begin{equation}
K_0 = \exp\!\left[-\frac{\kappa_b\,\delta t}{2}\,\hat n_b\right],
\label{eq:E0}
\end{equation}
where \(\kappa_b\) is the energy-decay rate of the \(b\) cavity. The time evolution under this Kraus operator to leading order in $\kappa_b \delta t$ is \cite{michael2016new}

\begin{equation}
\label{eq:no_jump_expand}
\begin{aligned}
\frac{K_0|\psi\rangle}{\bigl\|K_0|\psi\rangle\bigr\|}
&= \Bigl[\,1 + \frac{\kappa_b \delta t}{2}\,(\bar n_b - \hat n_b)\Bigr]|\psi\rangle
   + \mathcal{O}\!\left[(\kappa_b \delta t)^2\right] \\
&= u\Bigl(|0_{\mathrm L}'\rangle + \kappa_b \delta t\,|0_{\mathrm E}'\rangle\Bigr)
   + v\,|1_{\mathrm L}'\rangle
   + \mathcal{O}\!\left[(\kappa_b \delta t)^2\right].
\end{aligned}
\end{equation}
Notice that only \(|0_{\mathrm{L}}'\rangle\) acquires a component outside the code space, with the ``error word"
\begin{equation}
|0_{\mathrm E}'\rangle \;\equiv\;
  \frac{|0 \ 0 \ 1\rangle - |0 \ 4 \ 1\rangle}{\sqrt{2}},
\end{equation}
which is orthogonal to \( |0_{\mathrm{L}}'\rangle, |1_{\mathrm{L}}'\rangle\). Crucially, we observe that the no-jump back-action is equivalent to a small rotation within the subspace $\{|0_{\mathrm{L}}'\rangle,\,|0_{\mathrm{E}}'\rangle\}$ to first order in $\kappa_b\delta t$. This evolution can thus be corrected to first order by targeting the states 
\begin{align}
|\widetilde{\psi}_{\mathrm{tgt}}^{(k)}\rangle
=
\frac{
U_0|\psi_{\mathrm{tgt}}^{(k)}\rangle
}{
\|U_0|\psi_{\mathrm{tgt}}^{(k)}\rangle\|
}.
\end{align}
where we have defined the unitary
\cite{michael2016new}
\begin{equation}
\label{eq:U0}
\begin{aligned}
\ U_0=\;&
\sin(\kappa_b\delta t)\bigl(|0_{\text{L}}'\rangle\langle 0_{\text{E}}'| - |0_{\text{E}}'\rangle\langle 0_{\text{L}}'|\bigr)\\
&+\cos(\kappa_b\delta t)\bigl(|0_{\text{L}}'\rangle\langle 0_{\text{L}}'| + |0_{\text{E}}'\rangle\langle 0_{\text{E}}'|\bigr)\\
&+|1_{\text{L}}'\rangle\langle 1_{\text{L}}'|+ U_{\mathrm{res}},
\end{aligned}
\end{equation}
which effectively undoes the evolution caused by the no-jump back-action to leading order. The unitary $U_{\mathrm{res}}$ acts as the identity outside the relevant subspace. Thus, rather than targeting the ideal final states directly, we target states that are pre-rotated by $U_0$.

The GRAPE-optimized control fields for the conditional-pumping, conditional-routing, and superposition gates are shown in Fig.~\ref{fig:figure4}. As photon loss and relaxation during the gates are not correctable by the protocol, shorter gate durations are preferable. We therefore choose the shortest pulse durations for which the optimized control amplitudes remain below $5~\mathrm{MHz}$, a scale consistent with recent experimental implementations of parametrically activated cavity-cavity interactions \cite{lu2023high, chapman2022high}. Assuming that the durations of the pumping and reset operations (not shown here) are comparable to those of our optimized gates, the full protocol sequence can be completed in less than $10~\text{\textmu s}$. The GRAPE software we use is described in \citep{eickbusch2022fast}.

\section{Results}
\label{sec:results}

Here we present results from QuTiP \citep{johansson2013qutip} simulations of the protocol. We model three sources of infidelity: coherent control errors from the finite-fidelity optimized gates, cavity photon loss, and transmon relaxation and dephasing. State preparation and reset are treated as ideal operations, and infidelities associated with the coupler modes are not included. Unless otherwise stated, we simulate the cavities with $T_1^{\mathrm{cav}}=1~\mathrm{ms}$ and the transmons with $T_1^{\mathrm{tr}}=T_2^{\mathrm{tr}}=300~\mathrm{\mu s}$.

We use QuTiP's Lindblad master equation solver \texttt{mesolve} to simulate the protocol under open-system dynamics. The error-correction map is evaluated on the six cardinal logical states
$|0_{\mathrm{L}}\rangle$, $|1_{\mathrm{L}}\rangle$, $|+_{\mathrm{L}}\rangle$,
$|-_{\mathrm{L}}\rangle$, $|+i_{\mathrm{L}}\rangle$, and
$|-i_{\mathrm{L}}\rangle$, as well as their corresponding single-photon-loss states. For each logical input and its corresponding single-loss input, the target is the   restored logical state. We define the single-round correction fidelity as
\begin{equation}
F_{\psi}
=
\langle \Psi_{\mathrm{tgt}} |
\rho^{\mathrm{final}}
| \Psi_{\mathrm{tgt}} \rangle,
\label{eq:single-round-fidelity}
\end{equation}
where $\rho^{\mathrm{final}}$ is the final density matrix of the full system, and $|\Psi_{\mathrm{tgt}}\rangle$ is the target state. Averaging over the six cardinal logical inputs gives
$F_{\mathrm{card}}=0.992$, which represents the fidelity of the no-error branch. Averaging over the corresponding single-loss inputs gives
$F_{\mathrm{loss}}=0.990$, which represents the fidelity of the error branch.

\begin{figure}[!t]
    \centering
    \includegraphics[width=\linewidth]{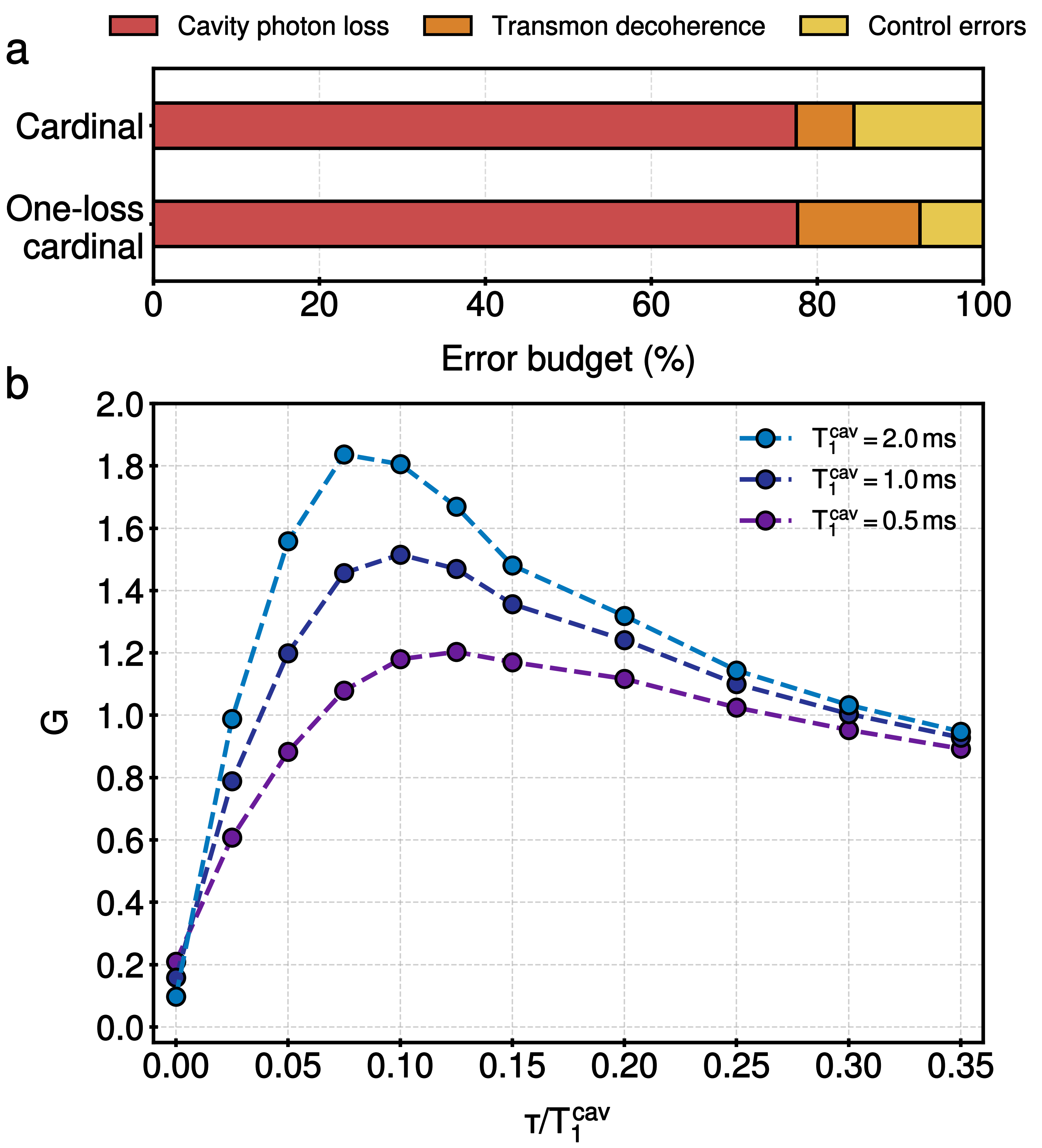}
    \caption{\textbf{Single-round error budget and repetitive-QEC performance.}
    \textbf{(a)} Error budget for a single application of the error-correction
    protocol, shown separately for the six cardinal logical states and their
    one-photon-loss counterparts. These simulations use
    $T_1^{\mathrm{cav}}=1~\mathrm{ms}$ for all three cavities and
    $T_1^{\mathrm{tr}}=T_2^{\mathrm{tr}}=300~\mathrm{\mu s}$ for all auxiliary
    transmons. The contributions from cavity photon loss, transmon decoherence,
    and coherent-control error are normalized to sum to $100\%$ in each bar and small nonadditive residuals are omitted.
    \textbf{(b)} QEC gain $G(\tau)$ extracted from logical process-fidelity decay
    under periodic application of the error-correction protocol with inter-cycle
    waiting time $\tau$. For nonzero $\tau$, an idle no-jump correction gate is
    applied before each recovery sequence to compensate the deterministic
    no-jump back-action accumulated during the waiting time. The cavity lifetime
    $T_1^{\mathrm{cav}}$ is varied as indicated, while the transmon coherence
    times are held fixed at
    $T_1^{\mathrm{tr}}=T_2^{\mathrm{tr}}=300~\mathrm{\mu s}$.
    The decay curves are fit over an observation window of
    $3T_1^{\mathrm{cav}}$, with elapsed time including both idle intervals
    and correction gates.
    For $T_1^{\mathrm{cav}}=0.5~\mathrm{ms}$, $1~\mathrm{ms}$, and
    $2~\mathrm{ms}$, the maximum gains are $G=1.20$, $1.52$, and $1.84$,
    respectively, occurring at
    $\tau/T_1^{\mathrm{cav}}=0.125$, $0.100$, and $0.075$.
    At shorter waiting times, the performance is constrained by uncorrectable
    photon losses during the protocol itself, whereas at longer waiting times
    it is limited by multiple photon losses accumulated between QEC cycles.}
\label{fig:figure5}
\end{figure}

We compute an error budget to identify the dominant sources of infidelity in a single application of the protocol, see Fig.~\hyperref[fig:figure5]{\ref*{fig:figure5}a}. For cavity and transmon decoherence, each contribution is obtained by comparing the full simulation with an otherwise identical simulation in which the corresponding collapse operators are removed. Denoting the full-simulation fidelity by $F_{\mathrm{all}}$ and the fidelity obtained after removing mechanism $x$ by $F_{-x}$, we define
\begin{equation}
\epsilon_x = F_{-x}-F_{\mathrm{all}}.
\end{equation}
The coherent-control contribution, $\epsilon_{\mathrm{ctrl}}$, is estimated from the closed-system infidelity of the implemented gate sequence.

For the cardinal states, the total single-round infidelity is
$\epsilon_{\mathrm{tot}}=8 \times10^{-3}$, with cavity photon loss, transmon decoherence, and coherent-control error contributing $77.5\%$, $7.0\%$, and $15.6\%$, respectively. For the one-loss cardinal states, the total infidelity is
$\epsilon_{\mathrm{tot}}=1\times10^{-2}$, with the corresponding contributions $77.6\%$, $14.8\%$, and $7.6\%$. The larger relative contribution from transmon decoherence in the one-loss branch arises because correction of a photon loss requires the protocol to pass through the single-cavity superposition gate, during which the relevant logical information is temporarily coupled to the auxiliary transmon. Nevertheless, in both branches the dominant limitation is cavity photon loss during the gate sequence. Transmon decoherence remains a subleading contribution, indicating that the protocol is not primarily limited by the auxiliary transmons.

To assess whether repeated correction extends the encoded lifetime, we simulate a memory experiment in which the protocol is applied periodically. Between correction cycles, the storage cavity evolves under photon loss for an idle time $\tau$. Even when no photon-loss event occurs during this interval, the conditional no-jump evolution produces a deterministic deformation of the encoded state, analogous to the effect discussed in Sec.~\ref{sec:gates}. To compensate this accumulated back-action, we apply a no-jump correction gate to the storage cavity immediately before each error-correction cycle. This gate is optimized with GRAPE using the same single-cavity dispersive Hamiltonian as the superposition gate in Eq.~\eqref{eq:9}, but applied to cavity $b$ rather than cavity $a$. For each value of $\tau$, the target unitary is chosen to invert the leading order no-jump deformation accumulated during the idle time, while preserving the relevant single-loss error states. Thus, one repeated-QEC cycle consists of idle evolution for time $\tau$, the corresponding no-jump correction gate, and the full measurement-free recovery protocol.

Note that the choice of $\tau$ must balance two competing effects. If the protocol is applied too frequently, a larger fraction of the photon losses occur during the error correction protocol, which cannot be corrected. If it is applied too infrequently, two or more photon losses can accumulate between correction cycles, which the lowest-order binomial code cannot correct. We determine the optimally balanced waiting time by simulating repeated QEC cycles for different values of $\tau$ and comparing the resulting logical lifetime to that of the unprotected Fock $\{|0\rangle,|1\rangle\}$  encoding.

For both the binomial and Fock encodings, we compute the logical
process fidelity with respect to the
identity channel as a function of the total elapsed time.
Denoting the effective logical channel on the storage cavity after time $t$ by
$\mathcal{E}_t$, we evaluate
\begin{equation}
F_{\rm e}(t)
=
\frac{1}{4}
\sum_{i,j=0}^{1}
\langle i_{\rm L}|
\mathcal{E}_t\!\left(|i_{\rm L}\rangle\langle j_{\rm L}|\right)
|j_{\rm L}\rangle .
\label{eq:entanglement_fidelity}
\end{equation}
Note that this quantity is distinct from the state fidelity in Eq.~(\ref{eq:single-round-fidelity}),
which is used only as a single-round diagnostic. For each waiting time $\tau$, we fit the simulated
entanglement-fidelity decay over an observation window of
$3T_1^{\mathrm{cav}}$ to a single-exponential form,
\begin{equation}
F_{\rm e}(t)=C+A e^{-t/T_{\rm eff}},
\end{equation}
using the same fitting convention for the binomial and Fock encodings. The
QEC gain is then defined as
\begin{equation}
G(\tau)\equiv
\frac{T_{\rm eff}^{\rm QEC}(\tau)}
{T_{\rm eff}^{\rm Fock}} .
\end{equation}
For a cavity lifetime $T_1^{\mathrm{cav}}=1~\mathrm{ms}$, the gain is maximized near $\tau/T_1^{\mathrm{cav}}=0.10$, where $G=1.52$. We repeat the simulation for cavity lifetimes $T_1^{\mathrm{cav}}=0.5~\mathrm{ms}$ and $T_1^{\mathrm{cav}}=2~\mathrm{ms}$, keeping
$T_1^{\rm tr}=T_2^{\rm tr}=300~\mathrm{\mu s}$ and using the same optimized
gate set. The maximum gains are $G=1.20$ at
$\tau/T_1^{\mathrm{cav}}=0.125$ for
$T_1^{\mathrm{cav}}=0.5~\mathrm{ms}$ and $G=1.84$ at
$\tau/T_1^{\mathrm{cav}}=0.075$ for
$T_1^{\mathrm{cav}}=2~\mathrm{ms}$. As the cavity lifetime increases, the optimal waiting time shifts to smaller $\tau/T_1^{\mathrm{cav}}$. This trend reflects the fixed duration of the gate sequence: for longer-lived cavities, the correction cycle occupies a smaller fraction of the cavity lifetime, allowing the protocol to be applied more frequently before photon loss during the gates becomes the dominant limitation. In all three cases, the repeated-QEC protocol exceeds break-even relative to the bare Fock encoding, see Fig.~\hyperref[fig:figure5]{\ref*{fig:figure5}b}. Repeating the analysis with a fixed number of correction cycles,
rather than an observation window of $3T_1^{\mathrm{cav}}$,
yields comparable peak gains and preserves both the optimal
waiting times on the sampled grid and the increase in peak
gain with cavity lifetime.

\renewcommand{\figurename}{FIG.}

\section{Conclusion}
In this work, we have proposed a measurement-free error-correction scheme for a binomial-code logical qubit encoded in a superconducting cavity, where entropy is removed using only unconditional resets. We modify and simplify our protocol compared to that proposed by Krastanov {\it et al.} \cite{krastanov2021room},
where the use of the lowest-order binomial code \cite{michael2016new} over the CLY code \cite{chuang1997bosonic} reduces the hardware requirement from six to three quantum modes. We specifically focus on the implementation of this protocol using superconducting circuits and cavities.
By leveraging the remarkable Hamiltonian engineering capabilities of cQED, we introduce one additional degree of freedom in gate design compared to the fixed \(\chi^{(2)}\) nonlinearity used in Ref.~\cite{krastanov2021room}. Our protocol is all-unitary, with the exception of the final unconditional reset gate. The strategy pursued here differs from dissipative autonomous QEC \citep{paz1998continuous, kapit2016hardware, kapit2017upside, lihm2018implementation, lescanne2020exponential}, which typically relies on complex bath engineering to non-unitarily drive error states into the code space. 

As we cannot correct photon losses occurring during the actual error-correction protocol, we design short pulse sequences to operate on timescales significantly faster than the cavity decay rate. In simulations, a sub-5\,\textmu s gate sequence yields correction fidelities of \(0.99\) across the cardinal (and corresponding error) states. For the parameters considered here, the dominant contribution to single-round QEC is cavity photon loss during the correction cycle, rather than decoherence from the ancilla transmons. This indicates that improving the cavity lifetime should directly increase the achievable QEC gain. Consistent with this expectation, our repeated-cycle simulations show that the maximum gain increases as the cavity lifetime is improved. For a cavity lifetime $T_1^{\mathrm{cav}} = 1$~ms, the optimal waiting time is
$\tau \approx 0.10\,T_1^{\mathrm{cav}}$, yielding a logical lifetime enhancement of $1.52$ relative to the bare Fock encoding and surpassing break-even performance. Our results support a practical, gate-based, measurement-free QEC approach that is ready for near-term implementation.

\section{Acknowledgments}
DKW and SMG acknowledge support from the Air Force Office of Scientific Research under award number FA9550-21-1-0209.  SMG also acknowledges support from the Army Research Office (ARO) under 
Grant Number W911NF-23-1-0051. The views and conclusions contained in this document are 
those of the authors and should not be interpreted as representing the official policies, 
either expressed or implied, of the Army Research Office (ARO), or the U.S.\ Government. 
The U.S.\ Government is authorized to reproduce and distribute reprints for Government purposes notwithstanding any copyright notation herein.

External interest disclosure: SMG is a consultant for D-Wave Systems, Inc. and a co-founder and equity holder in Toric Labs, Inc.

\bibliography{references}

@article{chapman2022high,
	title        = {High-On-Off-Ratio Beam-Splitter Interaction for Gates on Bosonically Encoded Qubits},
	author       = {Chapman, Benjamin J. and de Graaf, Stijn J. and Xue, Sophia H. and Zhang, Yaxing and Teoh, James and Curtis, Jacob C. and Tsunoda, Takahiro and Eickbusch, Alec and Read, Alexander P. and Koottandavida, Akshay and Mundhada, Shantanu O. and Frunzio, Luigi and Devoret, M. H. and Girvin, S. M. and Schoelkopf, R. J.},
	year         = 2023,
	month        = {Jun},
journal      = {PRX Quantum},
	publisher    = {American Physical Society},
	volume       = 4,
	pages        = {020355},
	doi          = {10.1103/PRXQuantum.4.020355},
	url          = {https://link.aps.org/doi/10.1103/PRXQuantum.4.020355},
	issue        = 2,
	numpages     = 30
}

@article{chou2023demonstrating,
	title        = {A superconducting dual-rail cavity qubit with erasure-detected logical measurements},
	author       = {Chou, Kevin S. and Shemma, Tali and McCarrick, Heather and Chien, Tzu-Chiao and Teoh, James D. and Winkel, Patrick and Anderson, Amos and Chen, Jonathan and Curtis, Jacob C. and de Graaf, Stijn J. and Garmon, John W. O. and Gudlewski, Benjamin and Kalfus, William D. and Keen, Trevor and Khedkar, Nishaad and Lei, Chan U. and Liu, Gangqiang and Lu, Pinlei and Lu, Yao and Maiti, Aniket and Mastalli-Kelly, Luke and Mehta, Nitish and Mundhada, Shantanu O. and Narla, Anirudh and Noh, Taewan and Tsunoda, Takahiro and Xue, Sophia H. and Yuan, Joseph O. and Frunzio, Luigi and Aumentado, José and Puri, Shruti and Girvin, Steven M. and Moseley, S. Harvey and Schoelkopf, Robert J.},
	year         = 2024,
	month        = sep,
	journal      = {Nat. Phys.},
	volume       = 20,
	number       = 9,
	pages        = {1454--1460},
	doi          = {10.1038/s41567-024-02539-4},
	issn         = {1745-2481},
	url          = {https://doi.org/10.1038/s41567-024-02539-4}
}

@article{sun_tracking_2014,
	title = {Tracking photon jumps with repeated quantum non-demolition parity measurements},
	volume = {511},
	issn = {1476-4687},
	url = {https://doi.org/10.1038/nature13436},
	doi = {10.1038/nature13436},
	number = {7510},
	journal = {Nature},
	author = {Sun, L. and Petrenko, A. and Leghtas, Z. and Vlastakis, B. and Kirchmair, G. and Sliwa, K. M. and Narla, A. and Hatridge, M. and Shankar, S. and Blumoff, J. and Frunzio, L. and Mirrahimi, M. and Devoret, M. H. and Schoelkopf, R. J.},
	month = jul,
	year = {2014},
	pages = {444--448},
}

@article{ni2023beating,
	title = {Beating the break-even point with a discrete-variable-encoded logical qubit},
	volume = {616},
	issn = {1476-4687},
	url = {https://doi.org/10.1038/s41586-023-05784-4},
	doi = {10.1038/s41586-023-05784-4},
	number = {7955},
	journal = {Nature},
	author = {Ni, Zhongchu and Li, Sai and Deng, Xiaowei and Cai, Yanyan and Zhang, Libo and Wang, Weiting and Yang, Zhen-Biao and Yu, Haifeng and Yan, Fei and Liu, Song and Zou, Chang-Ling and Sun, Luyan and Zheng, Shi-Biao and Xu, Yuan and Yu, Dapeng},
	month = apr,
	year = {2023},
	pages = {56--60},
}

@article{vrajitoarea2020quantum,
  title     = {Quantum control of an oscillator using a stimulated Josephson nonlinearity},
  author    = {Vrajitoarea, Andrei and Huang, Ziwen and Groszkowski, Peter and Koch, Jens and Houck, Andrew A.},
  journal   = {Nature Physics},
  volume    = {16},
  number    = {2},
  pages     = {211--217},
  year      = {2020},
  publisher = {Springer Nature},
  doi       = {10.1038/s41567-019-0703-5}
}

@article{fowler2012surface,
  title   = {Surface codes: Towards practical large-scale quantum computation},
  author  = {Fowler, Austin G. and Mariantoni, Matteo and Martinis, John M. and Cleland, Andrew N.},
  journal = {Physical Review A},
  volume  = {86},
  number  = {3},
  pages   = {032324},
  year    = {2012},
  publisher = {American Physical Society},
  doi     = {10.1103/PhysRevA.86.032324}
}

@article{krastanov2021room,
  title     = {Room-temperature photonic logical qubits via second-order nonlinearities},
  author    = {Krastanov, Stefan and Heuck, Mikkel and Shapiro, Jeffrey H. and Narang, Prineha and Englund, Dirk R. and Jacobs, Kurt},
  journal   = {Nature Communications},
  volume    = {12},
  number    = {1},
  pages     = {191},
  year      = {2021},
  publisher = {Springer Nature},
  doi       = {10.1038/s41467-020-20417-4}
}

@article{gertler2021protecting,
  title={Protecting a bosonic qubit with autonomous quantum error correction},
  author={Gertler, Jeffrey M and Baker, Brian and Li, Juliang and Shirol, Shruti and Koch, Jens and Wang, Chen},
  journal={Nature},
  volume={590},
  number={7845},
  pages={243--248},
  year={2021},
  publisher={Nature Publishing Group UK London}
}

@article{milul2023superconducting,
  title     = {Superconducting cavity qubit with tens of milliseconds single-photon coherence time},
  author    = {Milul, Ofir and Guttel, Barkay and Goldblatt, Uri and Hazanov, Sergey and Joshi, Lalit M. and Chausovsky, Daniel and Kahn, Nitzan and {\c{C}}ifty{\"u}rek, Engin and Lafont, Fabien and Rosenblum, Serge},
  journal   = {PRX Quantum},
  volume    = {4},
  number    = {3},
  pages     = {030336},
  year      = {2023},
  publisher = {American Physical Society},
  doi       = {10.1103/PRXQuantum.4.030336}
}

@article{michael2016new,
  title     = {New class of quantum error-correcting codes for a bosonic mode},
  author    = {Michael, Marios H. and Silveri, Matti and Brierley, R. T. and Albert, Victor V. and Salmilehto, Juha and Jiang, Liang and Girvin, Steven M.},
  journal   = {Physical Review X},
  volume    = {6},
  number    = {3},
  pages     = {031006},
  year      = {2016},
  publisher = {American Physical Society},
  doi       = {10.1103/PhysRevX.6.031006}
}

@article{frattini20173,
  title     = {3-wave mixing Josephson dipole element},
  author    = {Frattini, N. E. and Vool, U. and Shankar, S. and Narla, A. and Sliwa, K. M. and Devoret, M. H.},
  journal   = {Applied Physics Letters},
  volume    = {110},
  number    = {22},
  pages     = {222603},
  year      = {2017},
  publisher = {AIP Publishing},
  doi       = {10.1063/1.4984142}
}

@article{pietikainen2024strategies,
  title     = {Strategies and Trade-Offs for Controllability and Memory Time of Ultra-High-Quality Microwave Cavities in Circuit Quantum Electrodynamics},
  author    = {Pietik{\"a}inen, Iivari and {\v{C}}ernot{\'\i}k, Ond{\v{r}}ej and Eickbusch, Alec and Maiti, Aniket and Garmon, John W. O. and Filip, Radim and Girvin, Steven M.},
  journal   = {PRX Quantum},
  volume    = {5},
  issue     = {4},
  pages     = {040307},
  numpages  = {30},
  year      = {2024},
  month     = {Oct},
  publisher = {American Physical Society},
  doi       = {10.1103/PRXQuantum.5.040307},
  url       = {https://link.aps.org/doi/10.1103/PRXQuantum.5.040307}
}

@article{pietikainen2022controlled,
  title     = {Controlled beam splitter gate transparent to dominant ancilla errors},
  author    = {Pietik{\"a}inen, Iivari and {\v{C}}ernot{\'\i}k, Ond{\v{r}}ej and Puri, Shruti and Filip, Radim and Girvin, S. M.},
  journal   = {Quantum Science and Technology},
  volume    = {7},
  number    = {3},
  pages     = {035025},
  year      = {2022},
  publisher = {IOP Publishing},
  doi       = {10.1088/2058-9565/ac760a}
}

@article{niu2018qudit,
  title     = {Qudit-Basis Universal Quantum Computation Using {$\chi^{(2)}$} Interactions},
  author    = {Niu, Murphy Yuezhen and Chuang, Isaac L. and Shapiro, Jeffrey H.},
  journal   = {Physical Review Letters},
  volume    = {120},
  number    = {16},
  pages     = {160502},
  year      = {2018},
  publisher = {American Physical Society},
  doi       = {10.1103/PhysRevLett.120.160502}
}

@article{bartlett2002efficient,
  title     = {Efficient Classical Simulation of Continuous Variable Quantum Information Processes},
  author    = {Bartlett, Stephen D. and Sanders, Barry C. and Braunstein, Samuel L. and Nemoto, Kae},
  journal   = {Physical Review Letters},
  volume    = {88},
  number    = {9},
  pages     = {097904},
  year      = {2002},
  publisher = {American Physical Society},
  doi       = {10.1103/PhysRevLett.88.097904}
}

@article{jacobs2007engineering,
  title     = {Engineering Quantum States of a Nanoresonator via a Simple Auxiliary System},
  author    = {Jacobs, Kurt},
  journal   = {Physical Review Letters},
  volume    = {99},
  number    = {11},
  pages     = {117203},
  year      = {2007},
  publisher = {American Physical Society},
  doi       = {10.1103/PhysRevLett.99.117203}
}

@article{krastanov2015universal,
  title     = {Universal control of an oscillator with dispersive coupling to a qubit},
  author    = {Krastanov, Stefan and Albert, Victor V. and Shen, Chao and Zou, Chang-Ling and Heeres, Reinier W. and Vlastakis, Brian and Schoelkopf, Robert J. and Jiang, Liang},
  journal   = {Physical Review A},
  volume    = {92},
  number    = {4},
  pages     = {040303},
  year      = {2015},
  publisher = {American Physical Society},
  doi       = {10.1103/PhysRevA.92.040303}
}

@article{eickbusch2022fast,
  title     = {Fast universal control of an oscillator with weak dispersive coupling to a qubit},
  author    = {Eickbusch, Alec and Sivak, Volodymyr and Ding, Andy Z. and Elder, Salvatore S. and Jha, Shantanu R. and Venkatraman, Jayameenakshi and Royer, Baptiste and Girvin, Steven M. and Schoelkopf, Robert J. and Devoret, Michel H.},
  journal   = {Nature Physics},
  volume    = {18},
  number    = {12},
  pages     = {1464--1469},
  year      = {2022},
  publisher = {Springer Nature},
  doi       = {10.1038/s41567-022-01776-9}
}

@article{hu2019quantum,
  title     = {Quantum error correction and universal gate set operation on a binomial bosonic logical qubit},
  author    = {Hu, Ling and Ma, Yuwei and Cai, Weizhou and Mu, Xianghao and Xu, Yuan and Wang, Weiting and Wu, Yukai and Wang, Haiyan and Song, Y. P. and Zou, C.-L. and others},
  journal   = {Nature Physics},
  volume    = {15},
  number    = {5},
  pages     = {503--508},
  year      = {2019},
  publisher = {Springer Nature},
  doi       = {10.1038/s41567-018-0414-3}
}

@article{blais2021circuit,
  title     = {Circuit quantum electrodynamics},
  author    = {Blais, Alexandre and Grimsmo, Arne L. and Girvin, Steven M. and Wallraff, Andreas},
  journal   = {Reviews of Modern Physics},
  volume    = {93},
  number    = {2},
  pages     = {025005},
  year      = {2021},
  publisher = {American Physical Society},
  doi       = {10.1103/RevModPhys.93.025005}
}

@article{paz1998continuous,
  title         = {Continuous error correction},
  author        = {Paz, Juan Pablo and Zurek, Wojciech Hubert},
  journal       = {Proceedings of the Royal Society of London. Series A: Mathematical, Physical and Engineering Sciences},
  volume        = {454},
  number        = {1969},
  pages         = {355--364},
  year          = {1998},
  publisher     = {The Royal Society},
  doi           = {10.1098/rspa.1998.0165}
}

@article{kapit2016hardware,
  title     = {Hardware-Efficient and Fully Autonomous Quantum Error Correction in Superconducting Circuits},
  author    = {Kapit, Eliot},
  journal   = {Physical Review Letters},
  volume    = {116},
  number    = {15},
  pages     = {150501},
  year      = {2016},
  publisher = {American Physical Society},
  doi       = {10.1103/PhysRevLett.116.150501}
}

@article{kapit2017upside,
  title     = {The upside of noise: engineered dissipation as a resource in superconducting circuits},
  author    = {Kapit, Eliot},
  journal   = {Quantum Science and Technology},
  volume    = {2},
  number    = {3},
  pages     = {033002},
  year      = {2017},
  publisher = {IOP Publishing},
  doi       = {10.1088/2058-9565/aa7e5d}
}

@article{lihm2018implementation,
  title     = {Implementation-independent sufficient condition of the Knill-Laflamme type for the autonomous protection of logical qudits by strong engineered dissipation},
  author    = {Lihm, Jae-Mo and Noh, Kyungjoo and Fischer, Uwe R.},
  journal   = {Physical Review A},
  volume    = {98},
  number    = {1},
  pages     = {012317},
  year      = {2018},
  publisher = {American Physical Society},
  doi       = {10.1103/PhysRevA.98.012317}
}

@article{lescanne2020exponential,
  title     = {Exponential suppression of bit-flips in a qubit encoded in an oscillator},
  author    = {Lescanne, Rapha{\"e}l and Villiers, Marius and Peronnin, Th{\'e}au and Sarlette, Alain and Delbecq, Matthieu and Huard, Benjamin and Kontos, Takis and Mirrahimi, Mazyar and Leghtas, Zaki},
  journal   = {Nature Physics},
  volume    = {16},
  number    = {5},
  pages     = {509--513},
  year      = {2020},
  publisher = {Springer Nature},
  doi       = {10.1038/s41567-020-0824-x}
}

@article{ofek2016extending,
  title     = {Extending the lifetime of a quantum bit with error correction in superconducting circuits},
  author    = {Ofek, Nissim and Petrenko, Andrei and Heeres, Reinier and Reinhold, Philip and Leghtas, Zaki and Vlastakis, Brian and Liu, Yehan and Frunzio, Luigi and Girvin, Steven M. and Jiang, Liang and others},
  journal   = {Nature},
  volume    = {536},
  number    = {7617},
  pages     = {441--445},
  year      = {2016},
  publisher = {Springer Nature},
  doi       = {10.1038/nature18949}
}

@article{shor1995scheme,
  title     = {Scheme for reducing decoherence in quantum computer memory},
  author    = {Shor, Peter W.},
  journal   = {Physical Review A},
  volume    = {52},
  number    = {4},
  pages     = {R2493--R2496},
  year      = {1995},
  publisher = {American Physical Society},
  doi       = {10.1103/PhysRevA.52.R2493}
}

@article{steane1996error,
  title     = {Error Correcting Codes in Quantum Theory},
  author    = {Steane, Andrew M.},
  journal   = {Physical Review Letters},
  volume    = {77},
  number    = {5},
  pages     = {793--797},
  year      = {1996},
  publisher = {American Physical Society},
  doi       = {10.1103/PhysRevLett.77.793}
}

@article{bombin2006topological,
  title     = {Topological quantum distillation},
  author    = {Bombin, Hector and Martin-Delgado, Miguel Angel},
  journal   = {Physical Review Letters},
  volume    = {97},
  number    = {18},
  pages     = {180501},
  year      = {2006},
  publisher = {American Physical Society},
  doi       = {10.1103/PhysRevLett.97.180501}
}

@article{koch2007charge,
  title     = {Charge-insensitive qubit design derived from the Cooper pair box},
  author    = {Koch, Jens and Yu, Terri M. and Gambetta, Jay and Houck, Andrew A. and Schuster, David I. and Majer, Johannes and Blais, Alexandre and Devoret, Michel H. and Girvin, Steven M. and Schoelkopf, Robert J.},
  journal   = {Physical Review A},
  volume    = {76},
  number    = {4},
  pages     = {042319},
  year      = {2007},
  publisher = {American Physical Society},
  doi       = {10.1103/PhysRevA.76.042319}
}

@article{gao2019entanglement,
  title     = {Entanglement of bosonic modes through an engineered exchange interaction},
  author    = {Gao, Yvonne Y. and Lester, Brian J. and Chou, Kevin S. and Frunzio, Luigi and Devoret, Michel H. and Jiang, Liang and Girvin, S. M. and Schoelkopf, Robert J.},
  journal   = {Nature},
  volume    = {566},
  number    = {7745},
  pages     = {509--512},
  year      = {2019},
  publisher = {Springer Nature},
  doi       = {10.1038/s41586-019-0970-4}
}

@article{lu2023high,
  title     = {High-fidelity parametric beamsplitting with a parity-protected converter},
  author    = {Lu, Yao and Maiti, Aniket and Garmon, John W. O. and Ganjam, Suhas and Zhang, Yaxing and Claes, Jahan and Frunzio, Luigi and Girvin, Steven M. and Schoelkopf, Robert J.},
  journal   = {Nature Communications},
  volume    = {14},
  number    = {1},
  pages     = {5767},
  year      = {2023},
  publisher = {Springer Nature},
  doi       = {10.1038/s41467-023-41104-0}
}

@article{khaneja2005optimal,
  title     = {Optimal control of coupled spin dynamics: design of NMR pulse sequences by gradient ascent algorithms},
  author    = {Khaneja, Navin and Reiss, Timo and Kehlet, Cindie and Schulte-Herbr{\"u}ggen, Thomas and Glaser, Steffen J.},
  journal   = {Journal of Magnetic Resonance},
  volume    = {172},
  number    = {2},
  pages     = {296--305},
  year      = {2005},
  publisher = {Elsevier},
  doi       = {10.1016/j.jmr.2004.11.004}
}

@article{heeres2015cavity,
  title     = {Cavity state manipulation using photon-number selective phase gates},
  author    = {Heeres, Reinier W. and Vlastakis, Brian and Holland, Eric and Krastanov, Stefan and Albert, Victor V. and Frunzio, Luigi and Jiang, Liang and Schoelkopf, Robert J.},
  journal   = {Physical Review Letters},
  volume    = {115},
  number    = {13},
  pages     = {137002},
  year      = {2015},
  publisher = {American Physical Society},
  doi       = {10.1103/PhysRevLett.115.137002}
}

@article{reagor2016quantum,
  title     = {Quantum memory with millisecond coherence in circuit QED},
  author    = {Reagor, Matthew and Pfaff, Wolfgang and Axline, Christopher and Heeres, Reinier W. and Ofek, Nissim and Sliwa, Katrina and Holland, Eric and Wang, Chen and Blumoff, Jacob and Chou, Kevin and others},
  journal   = {Physical Review B},
  volume    = {94},
  number    = {1},
  pages     = {014506},
  year      = {2016},
  publisher = {American Physical Society},
  doi       = {10.1103/PhysRevB.94.014506}
}

@article{ganjam2024surpassing,
  title     = {Surpassing millisecond coherence in on chip superconducting quantum memories by optimizing materials and circuit design},
  author    = {Ganjam, Suhas and Wang, Yanhao and Lu, Yao and Banerjee, Archan and Lei, Chan U and Krayzman, Lev and Kisslinger, Kim and Zhou, Chenyu and Li, Ruoshui and Jia, Yichen and others},
  journal   = {Nature Communications},
  volume    = {15},
  number    = {1},
  pages     = {3687},
  year      = {2024},
  publisher = {Springer Nature},
  doi       = {10.1038/s41467-024-47857-6}
}

@misc{KnillLaflamme1996concatenated,
  title         = {Concatenated Quantum Codes},
  author        = {Knill, Emanuel and Laflamme, Raymond},
  year          = {1996},
  eprint        = {quant-ph/9608012},
  archivePrefix = {arXiv},
  primaryClass  = {quant-ph},
  url           = {https://arxiv.org/abs/quant-ph/9608012}
}

@article{google2023suppressing,
	title = {Suppressing quantum errors by scaling a surface code logical qubit},
	volume = {614},
	issn = {1476-4687},
	url = {https://doi.org/10.1038/s41586-022-05434-1},
	doi = {10.1038/s41586-022-05434-1},
	number = {7949},
	journal = {Nature},
	author = {Acharya, Rajeev and Aleiner, Igor and Allen, Richard and Andersen, Trond I. and Ansmann, Markus and Arute, Frank and Arya, Kunal and Asfaw, Abraham and Atalaya, Juan and Babbush, Ryan and Bacon, Dave and Bardin, Joseph C. and Basso, Joao and Bengtsson, Andreas and Boixo, Sergio and Bortoli, Gina and Bourassa, Alexandre and Bovaird, Jenna and Brill, Leon and Broughton, Michael and Buckley, Bob B. and Buell, David A. and Burger, Tim and Burkett, Brian and Bushnell, Nicholas and Chen, Yu and Chen, Zijun and Chiaro, Ben and Cogan, Josh and Collins, Roberto and Conner, Paul and Courtney, William and Crook, Alexander L. and Curtin, Ben and Debroy, Dripto M. and Del Toro Barba, Alexander and Demura, Sean and Dunsworth, Andrew and Eppens, Daniel and Erickson, Catherine and Faoro, Lara and Farhi, Edward and Fatemi, Reza and Flores Burgos, Leslie and Forati, Ebrahim and Fowler, Austin G. and Foxen, Brooks and Giang, William and Gidney, Craig and Gilboa, Dar and Giustina, Marissa and Grajales Dau, Alejandro and Gross, Jonathan A. and Habegger, Steve and Hamilton, Michael C. and Harrigan, Matthew P. and Harrington, Sean D. and Higgott, Oscar and Hilton, Jeremy and Hoffmann, Markus and Hong, Sabrina and Huang, Trent and Huff, Ashley and Huggins, William J. and Ioffe, Lev B. and Isakov, Sergei V. and Iveland, Justin and Jeffrey, Evan and Jiang, Zhang and Jones, Cody and Juhas, Pavol and Kafri, Dvir and Kechedzhi, Kostyantyn and Kelly, Julian and Khattar, Tanuj and Khezri, Mostafa and Kieferová, Mária and Kim, Seon and Kitaev, Alexei and Klimov, Paul V. and Klots, Andrey R. and Korotkov, Alexander N. and Kostritsa, Fedor and Kreikebaum, John Mark and Landhuis, David and Laptev, Pavel and Lau, Kim-Ming and Laws, Lily and Lee, Joonho and Lee, Kenny and Lester, Brian J. and Lill, Alexander and Liu, Wayne and Locharla, Aditya and Lucero, Erik and Malone, Fionn D. and Marshall, Jeffrey and Martin, Orion and McClean, Jarrod R. and McCourt, Trevor and McEwen, Matt and Megrant, Anthony and Meurer Costa, Bernardo and Mi, Xiao and Miao, Kevin C. and Mohseni, Masoud and Montazeri, Shirin and Morvan, Alexis and Mount, Emily and Mruczkiewicz, Wojciech and Naaman, Ofer and Neeley, Matthew and Neill, Charles and Nersisyan, Ani and Neven, Hartmut and Newman, Michael and Ng, Jiun How and Nguyen, Anthony and Nguyen, Murray and Niu, Murphy Yuezhen and O’Brien, Thomas E. and Opremcak, Alex and Platt, John and Petukhov, Andre and Potter, Rebecca and Pryadko, Leonid P. and Quintana, Chris and Roushan, Pedram and Rubin, Nicholas C. and Saei, Negar and Sank, Daniel and Sankaragomathi, Kannan and Satzinger, Kevin J. and Schurkus, Henry F. and Schuster, Christopher and Shearn, Michael J. and Shorter, Aaron and Shvarts, Vladimir and Skruzny, Jindra and Smelyanskiy, Vadim and Smith, W. Clarke and Sterling, George and Strain, Doug and Szalay, Marco and Torres, Alfredo and Vidal, Guifre and Villalonga, Benjamin and Vollgraff Heidweiller, Catherine and White, Theodore and Xing, Cheng and Yao, Z. Jamie and Yeh, Ping and Yoo, Juhwan and Young, Grayson and Zalcman, Adam and Zhang, Yaxing and Zhu, Ningfeng and {Google Quantum AI}},
	month = feb,
	year = {2023},
	pages = {676--681},
}

@article{gao2018programmable,
  title     = {Programmable interference between two microwave quantum memories},
  author    = {Gao, Yvonne Y. and Lester, Brian J. and Zhang, Yaxing and Wang, Chen and Rosenblum, Serge and Frunzio, Luigi and Jiang, Liang and Girvin, S. M. and Schoelkopf, Robert J.},
  journal   = {Physical Review X},
  volume    = {8},
  number    = {2},
  pages     = {021073},
  year      = {2018},
  publisher = {American Physical Society},
  doi       = {10.1103/PhysRevX.8.021073}
}

@article{axline2018demand,
  title     = {On-demand quantum state transfer and entanglement between remote microwave cavity memories},
  author    = {Axline, Christopher J. and Burkhart, Luke D. and Pfaff, Wolfgang and Zhang, Mengzhen and Chou, Kevin and Campagne-Ibarcq, Philippe and Reinhold, Philip and Frunzio, Luigi and Girvin, S. M. and Jiang, Liang and others},
  journal   = {Nature Physics},
  volume    = {14},
  number    = {7},
  pages     = {705--710},
  year      = {2018},
  publisher = {Springer Nature},
  doi       = {10.1038/s41567-018-0115-y}
}

@article{google2025quantum,
	title = {Quantum error correction below the surface code threshold},
	volume = {638},
	issn = {1476-4687},
	url = {https://doi.org/10.1038/s41586-024-08449-y},
	doi = {10.1038/s41586-024-08449-y},
	number = {8052},
	journal = {Nature},
	author = {Acharya, Rajeev and Abanin, Dmitry A. and Aghababaie-Beni, Laleh and Aleiner, Igor and Andersen, Trond I. and Ansmann, Markus and Arute, Frank and Arya, Kunal and Asfaw, Abraham and Astrakhantsev, Nikita and Atalaya, Juan and Babbush, Ryan and Bacon, Dave and Ballard, Brian and Bardin, Joseph C. and Bausch, Johannes and Bengtsson, Andreas and Bilmes, Alexander and Blackwell, Sam and Boixo, Sergio and Bortoli, Gina and Bourassa, Alexandre and Bovaird, Jenna and Brill, Leon and Broughton, Michael and Browne, David A. and Buchea, Brett and Buckley, Bob B. and Buell, David A. and Burger, Tim and Burkett, Brian and Bushnell, Nicholas and Cabrera, Anthony and Campero, Juan and Chang, Hung-Shen and Chen, Yu and Chen, Zijun and Chiaro, Ben and Chik, Desmond and Chou, Charina and Claes, Jahan and Cleland, Agnetta Y. and Cogan, Josh and Collins, Roberto and Conner, Paul and Courtney, William and Crook, Alexander L. and Curtin, Ben and Das, Sayan and Davies, Alex and De Lorenzo, Laura and Debroy, Dripto M. and Demura, Sean and Devoret, Michel and Di Paolo, Agustin and Donohoe, Paul and Drozdov, Ilya and Dunsworth, Andrew and Earle, Clint and Edlich, Thomas and Eickbusch, Alec and Elbag, Aviv Moshe and Elzouka, Mahmoud and Erickson, Catherine and Faoro, Lara and Farhi, Edward and Ferreira, Vinicius S. and Burgos, Leslie Flores and Forati, Ebrahim and Fowler, Austin G. and Foxen, Brooks and Ganjam, Suhas and Garcia, Gonzalo and Gasca, Robert and Genois, {\'E}lie and Giang, William and Gidney, Craig and Gilboa, Dar and Gosula, Raja and {Grajales Dau}, Alejandro and Graumann, Dietrich and Greene, Alex and Gross, Jonathan A. and Habegger, Steve and Hall, John and Hamilton, Michael C. and Hansen, Monica and Harrigan, Matthew P. and Harrington, Sean D. and Heras, Francisco J. H. and Heslin, Stephen and Heu, Paula and Higgott, Oscar and Hill, Gordon and Hilton, Jeremy and Holland, George and Hong, Sabrina and Huang, Hsin-Yuan and Huff, Ashley and Huggins, William J. and Ioffe, Lev B. and Isakov, Sergei V. and Iveland, Justin and Jeffrey, Evan and Jiang, Zhang and Jones, Cody and Jordan, Stephen and Joshi, Chaitali and Juhas, Pavol and Kafri, Dvir and Kang, Hui and Karamlou, Amir H. and Kechedzhi, Kostyantyn and Kelly, Julian and Khaire, Trupti and Khattar, Tanuj and Khezri, Mostafa and Kim, Seon and Klimov, Paul V. and Klots, Andrey R. and Kobrin, Bryce and Kohli, Pushmeet and Korotkov, Alexander N. and Kostritsa, Fedor and Kothari, Robin and Kozlovskii, Borislav and Kreikebaum, John Mark and Kurilovich, Vladislav D. and Lacroix, Nathan and Landhuis, David and Lange-Dei, Tiano and Langley, Brandon W. and Laptev, Pavel and Lau, Kim-Ming and Le Guevel, Loïck and Ledford, Justin and Lee, Joonho and Lee, Kenny and Lensky, Yuri D. and Leon, Shannon and Lester, Brian J. and Li, Wing Yan and Li, Yin and Lill, Alexander T. and Liu, Wayne and Livingston, William P. and Locharla, Aditya and Lucero, Erik and Lundahl, Daniel and Lunt, Aaron and Madhuk, Sid and Malone, Fionn D. and Maloney, Ashley and Mandrà, Salvatore and Manyika, James and Martin, Leigh S. and Martin, Orion and Martin, Steven and Maxfield, Cameron and McClean, Jarrod R. and McEwen, Matt and Meeks, Seneca and Megrant, Anthony and Mi, Xiao and Miao, Kevin C. and Mieszala, Amanda and Molavi, Reza and Molina, Sebastian and Montazeri, Shirin and Morvan, Alexis and Movassagh, Ramis and Mruczkiewicz, Wojciech and Naaman, Ofer and Neeley, Matthew and Neill, Charles and Nersisyan, Ani and Neven, Hartmut and Newman, Michael and Ng, Jiun How and Nguyen, Anthony and Nguyen, Murray and Ni, Chia-Hung and Niu, Murphy Yuezhen and O’Brien, Thomas E. and Oliver, William D. and Opremcak, Alex and Ottosson, Kristoffer and Petukhov, Andre and Pizzuto, Alex and Platt, John and Potter, Rebecca and Pritchard, Orion and Pryadko, Leonid P. and Quintana, Chris and Ramachandran, Ganesh and Reagor, Matthew J. and Redding, John and Rhodes, David M. and Roberts, Gabrielle and Rosenberg, Eliott and Rosenfeld, Emma and Roushan, Pedram and Rubin, Nicholas C. and Saei, Negar and Sank, Daniel and Sankaragomathi, Kannan and Satzinger, Kevin J. and Schurkus, Henry F. and Schuster, Christopher and Senior, Andrew W. and Shearn, Michael J. and Shorter, Aaron and Shutty, Noah and Shvarts, Vladimir and Singh, Shraddha and Sivak, Volodymyr and Skruzny, Jindra and Small, Spencer and Smelyanskiy, Vadim and Smith, W. Clarke and Somma, Rolando D. and Springer, Sofia and Sterling, George and Strain, Doug and Suchard, Jordan and Szasz, Aaron and Sztein, Alex and Thor, Douglas and Torres, Alfredo and Torunbalci, M. Mert and Vaishnav, Abeer and Vargas, Justin and Vdovichev, Sergey and Vidal, Guifre and Villalonga, Benjamin and Heidweiller, Catherine Vollgraff and Waltman, Steven and Wang, Shannon X. and Ware, Brayden and Weber, Kate and Weidel, Travis and White, Theodore and Wong, Kristi and Woo, Bryan W. K. and Xing, Cheng and Yao, Z. Jamie and Yeh, Ping and Ying, Bicheng and Yoo, Juhwan and Yosri, Noureldin and Young, Grayson and Zalcman, Adam and Zhang, Yaxing and Zhu, Ningfeng and Zobrist, Nicholas and {Google Quantum AI and Collaborators}},
	month = feb,
	year = {2025},
	pages = {920--926},
}

@article{chuang1997bosonic,
  title     = {Bosonic quantum codes for amplitude damping},
  author    = {Chuang, Isaac L. and Leung, Debbie W. and Yamamoto, Yoshihisa},
  journal   = {Physical Review A},
  volume    = {56},
  number    = {2},
  pages     = {1114--1125},
  year      = {1997},
  publisher = {American Physical Society},
  doi       = {10.1103/PhysRevA.56.1114}
}

@article{ni2025autonomous,
  title   = {Extending qubit lifetime beyond break-even and enhancing quantum sensing by autonomous quantum error correction},
  author  = {Ni, Zhongchu and Hu, Ling and Cai, Yanyan and
             Zhang, Libo and Mai, Jiasheng and Deng, Xiaowei and
             Zheng, Pan and Liu, Song and Zheng, Shi-Biao and
             Xu, Yuan and Yu, Dapeng},
  journal = {Nature Physics},
  year    = {2026},
  month   = nov,
  doi     = {10.1038/s41567-026-03493-z},
  url     = {https://doi.org/10.1038/s41567-026-03493-z}
}

@article{shirol2025passive,
  title   = {Passive Quantum Error Correction of Photon
             Loss at Breakeven},
  author  = {Shirol, Shruti and van Geldern, Sean
             and Xi, Hanzhe and Wang, Chen},
  journal = {Physical Review X},
  volume  = {16},
  pages   = {021042},
  year    = {2026},
  doi     = {10.1103/nvbm-97vs}
}

@article{bland2025millisecond,
  title     = {Millisecond lifetimes and coherence times in 2D transmon qubits},
  author    = {Bland, Matthew P. and Bahrami, Faranak and Martinez, Jeronimo G. C. and Prestegaard, Paal H. and Smitham, Basil M. and Joshi, Atharv and Hedrick, Elizabeth and Kumar, Shashwat and Yang, Ambrose and Pakpour-Tabrizi, Alexander C. and others},
  journal   = {Nature},
  volume    = {647},
  number    = {8089},
  pages     = {343--348},
  year      = {2025},
  publisher = {Springer Nature},
  doi       = {10.1038/s41586-025-09687-4}
}

@article{johansson2013qutip,
  title   = {QuTiP 2: A Python framework for the dynamics of open quantum systems},
  journal = {Computer Physics Communications},
  volume  = {184},
  number  = {4},
  pages   = {1234-1240},
  year    = {2013},
  issn    = {0010-4655},
  doi     = {10.1016/j.cpc.2012.11.019},
  author  = {J.R. Johansson and P.D. Nation and Franco Nori}
}

@misc{cai2026error,
  title         = {Error-detectable Universal Control for High-Gain Bosonic Quantum Error Correction},
  author        = {Cai, Weizhou and Chen, Zi-Jie and Li, Ming and Jie, Qing-Xuan and Zou, Xu-Bo and Guo, Guang-Can and Sun, Luyan and Zou, Chang-Ling},
  year          = {2026},
  eprint        = {2601.21838},
  archivePrefix = {arXiv},
  primaryClass  = {quant-ph},
  doi           = {10.48550/arXiv.2601.21838},
  url           = {https://arxiv.org/abs/2601.21838}
}

@article{cohen2014dissipation,
  title   = {Dissipation-induced continuous quantum error
             correction for superconducting circuits},
  author  = {Cohen, Joachim and Mirrahimi, Mazyar},
  journal = {Physical Review A},
  volume  = {90},
  pages   = {062344},
  year    = {2014},
  doi     = {10.1103/PhysRevA.90.062344}
}

@article{sun2026extending,
  title   = {Extension of coherence time beyond breakeven point using only drives and dissipation},
  author  = {Sun, Lida and Xu, Yifang and Zhou, Yilong and
             Hua, Ziyue and Wang, Weiting and Zhou, Jie and
             Chen, Zi-Jie and de Paula, Lui Zuccherelli and
             Jie, Qing-Xuan and Xue, Guangming and Yu, Haifeng and
             Cai, Weizhou and Zou, Chang-Ling and Sun, Luyan},
  journal = {Nature Physics},
  year    = {2026},
  month   = nov,
  doi     = {10.1038/s41567-026-03528-5},
  url     = {https://doi.org/10.1038/s41567-026-03528-5}
}
\end{document}